\input harvmac
\noblackbox

\let\includefigures=\iftrue
\let\useblackboard=\iftrue
\newfam\black

\includefigures
\message{If you do not have epsf.tex (to include figures),}
\message{change the option at the top of the tex file.}
\input epsf
\def\figin{\epsfcheck\figin}\def\figins{\epsfcheck\figins}
\def\epsfcheck{\ifx\epsfbox\UnDeFiNeD
\message{(NO epsf.tex, FIGURES WILL BE IGNORED)}
\gdef\figin##1{\vskip2in}\gdef\figins##1{\hskip.5in}
\else\message{(FIGURES WILL BE INCLUDED)}%
\gdef\figin##1{##1}\gdef\figins##1{##1}\fi}
\def\DefWarn#1{}
\def\figinsert{\goodbreak\midinsert}
\def\ifig#1#2#3{\DefWarn#1\xdef#1{fig.~\the\figno}
\writedef{#1\leftbracket fig.\noexpand~\the\figno}%
\figinsert\figin{\centerline{#3}}\medskip\centerline{\vbox{
\baselineskip12pt\advance\hsize by -1truein
\noindent\footnotefont{\bf Fig.~\the\figno:} #2}}
\bigskip\endinsert\global\advance\figno by1}
\else
\def\ifig#1#2#3{\xdef#1{fig.~\the\figno}
\writedef{#1\leftbracket fig.\noexpand~\the\figno}%
\global\advance\figno by1}
\fi
%

\useblackboard
\message{If you do not have msbm (blackboard bold) fonts,}
\message{change the option at the top of the tex file.}
\font\blackboard=msbm10 scaled \magstep1
\font\blackboards=msbm7
\font\blackboardss=msbm5
\textfont\black=\blackboard
\scriptfont\black=\blackboards
\scriptscriptfont\black=\blackboardss

\else

\fi
%

\def\yboxit#1#2{\vbox{\hrule height #1 \hbox{\vrule width #1
\vbox{#2}\vrule width #1 }\hrule height #1 }}
\def\fillbox#1{\hbox to #1{\vbox to #1{\vfil}\hfil}}
\def\ybox{{\lower 1.3pt \yboxit{0.4pt}{\fillbox{8pt}}\hskip-0.2pt}}
%
%


\def\comments#1{}

\def\half{{1\over 2}}

\def\vev#1{\langle{#1}\rangle}

\def\CC{{\cal C}}

\def\CE{{\cal E}}
\def\CF{{\cal F}}

\def\CT{{\cal T}}

\def\CN{{\cal N}}
\def\CO{{\cal O}}


\def\II{\relax{I\kern-.10em I}}

\def\cascade{{\cal A}}
%

\def\IZ{\relax{\rm Z\kern-.34em Z}}
\def\IB{\relax{\rm I\kern-.18em B}}
\def\IC{{\relax\hbox{$\inbar\kern-.3em{\rm C}$}}}
\def\ID{\relax{\rm I\kern-.18em D}}
\def\IE{\relax{\rm I\kern-.18em E}}
\def\IF{\relax{\rm I\kern-.18em F}}
\def\IG{\relax\hbox{$\inbar\kern-.3em{\rm G}$}}
\def\IGa{\relax\hbox{${\rm I}\kern-.18em\Gamma$}}
\def\IH{\relax{\rm I\kern-.18em H}}
\def\II{\relax{\rm I\kern-.18em I}}
\def\IK{\relax{\rm I\kern-.18em K}}
\def\IP{\relax{\rm I\kern-.18em P}}

%

\def\inbar{\,\vrule height1.5ex width.4pt depth0pt}

\def\IR{\relax{\rm I\kern-.18em R}}


\def\simlt{\hskip0.05in\relax{ 
\raise3.0pt\hbox{ $<$
{\lower5.0pt\hbox{\kern-1.05em $\sim$}} }} \hskip0.05in}
\def\simgt{\hskip0.05in\relax{ 
\raise3.0pt\hbox{ $>$
{\lower5.0pt\hbox{\kern-1.05em $\sim$}}}} \hskip0.05in}

%


%

\def\lp10{\ell_p^{10}}
\def\lp11{\ell_p^{11}}
\def\R11{R_{11}}

\def\frac#1#2{{#1 \over #2}}



\newdimen\tableauside\tableauside=1.0ex
\newdimen\tableaurule\tableaurule=0.4pt
\newdimen\tableaustep
\def\phantomhrule#1{\hbox{\vbox to0pt{\hrule height\tableaurule width#1\vss}}}
\def\phantomvrule#1{\vbox{\hbox to0pt{\vrule width\tableaurule height#1\hss}}}
\def\sqr{\vbox{%
  \phantomhrule\tableaustep
  \hbox{\phantomvrule\tableaustep\kern\tableaustep\phantomvrule\tableaustep}%
  \hbox{\vbox{\phantomhrule\tableauside}\kern-\tableaurule}}}
\def\squares#1{\hbox{\count0=#1\noindent\loop\sqr
  \advance\count0 by-1 \ifnum\count0>0\repeat}}
\def\tableau#1{\vcenter{\offinterlineskip
  \tableaustep=\tableauside\advance\tableaustep by-\tableaurule
  \kern\normallineskip\hbox
    {\kern\normallineskip\vbox
      {\gettableau#1 0 }%
     \kern\normallineskip\kern\tableaurule}%
  \kern\normallineskip\kern\tableaurule}}
\def\gettableau#1 {\ifnum#1=0\let\next=\null\else
  \squares{#1}\let\next=\gettableau\fi\next}

\tableauside=1.0ex
\tableaurule=0.4pt


 %
 %
 \def\eqnn#1{\xdef #1{(\secsym\the\meqno)}\writedef{#1\leftbracket#1}%
 \global\advance\meqno by1\wrlabeL#1}
 \def\eqna#1{\xdef #1##1{\hbox{$(\secsym\the\meqno##1)$}}
 \writedef{#1\numbersign1\leftbracket#1{\numbersign1}}%
 \global\advance\meqno by1\wrlabeL{#1$\{\}$}}
 \def\eqn#1#2{\xdef #1{(\secsym\the\meqno)}\writedef{#1\leftbracket#1}%
 \global\advance\meqno by1$$#2\eqno#1\eqlabeL#1$$}

\global\newcount\itemno \global\itemno=0

\def\itemaut#1{\global\advance\itemno by1\noindent\item{\the\itemno.}#1}



\def\({\left(}
\def\){\right)}
\def\[{\left[}
\def\]{\right]}
\def\<{\langle}
\def\>{\rangle}
\def\half{{1\over 2}}

\def\eg{{\it e.g.}}
\def\ie{{\it i.e.}}

\def\etal{{\it et. al.}}
\hyphenation{Di-men-sion-al}



\lref\NishidaPJ{
  Y.~Nishida and D.~T.~Son,
 ``Nonrelativistic conformal field theories,''
  Phys.\ Rev.\  D {\bf 76}, 086004 (2007)
  [arXiv:0706.3746 [hep-th]].
}

\lref\MehenND{
  T.~Mehen, I.~W.~Stewart and M.~B.~Wise,
  ``Conformal invariance for non-relativistic field theory,''
  Phys.\ Lett.\  B {\bf 474}, 145 (2000)
  [arXiv:hep-th/9910025].
}

\lref\cascade{
M.~Chertkov, I.~Kolokolov, and M.~Vergassola,
``Inverse versus Direct Cascades in Turbulent Advection,''
Phys.\ Rev.\ Lett.\ {\bf 80}, 512 (1998).}

\lref\sachdev{
S.~Sachdev, {\it Quantum Phase Transitions}, Cambridge, 1999.}

\lref\bulkref{
HHH, Kovtun-Sachdev..., Karch..., Liu, Rajagopal, Wiedemann
}

\lref\zw{K.~M.~O'Hara \etal,
  ``Observation of a strongly interacting degenerate Fermi gas of atoms,''
  Science {\bf 298}, 2179 (2002); C. A. Regal, M. Greiner, and D. S. Jin, 
``Observation of Resonance Condensation of Fermionic Atom Pairs,"
Phys.\ Rev.\ Lett.\ {\bf 92}, 040403 (2004); M.~Bartenstein \etal,
  ``Crossover from a molecular Bose-Einstein condensate to a
  degenerate Fermi gas,''
  Phys.\ Rev.\ Lett.\ {\bf 92}, 120401 (2004); M.~Zwierlein {\it et.\ al.}, 
``Condensation of Pairs of Fermionic Atoms Near a Feshbach Resonance,''
Phys.\ Rev.\ Lett.\ {\bf 92}, 120403 (2004);
 J.~Kinast \etal,
  ``Evidence for superfluidity in a resonantly interacting Fermi gas,''
  Phys.\ Rev.\ Lett.\ {\bf 92}, 150402 (2004);
  T.~Bourdel \etal,
  ``Experimental study of the BEC-BCS crossover region in lithium 6,''
  Phys.\ Rev.\ Lett.\ {\bf 93}, 050401 (2004).
}

\lref\LukierskiXY{
  J.~Lukierski, P.~C.~Stichel and W.~J.~Zakrzewski,
  ``Exotic Galilean conformal symmetry and its dynamical realisations,''
  Phys.\ Lett.\  A {\bf 357}, 1 (2006)
  [arXiv:hep-th/0511259].
}
\lref\ArdonneWA{
  E.~Ardonne, P.~Fendley and E.~Fradkin,
 ``Topological order and conformal quantum critical points,''
  Annals Phys.\  {\bf 310}, 493 (2004)
  [arXiv:cond-mat/0311466].
}

\lref\KLM{
S.\ Kachru, X.\ Liu, and M.\ Mulligan, 
``Holography and dynamical critical phenomena,"
to appear.}

\lref\damson{D.~Son, arXiv:0804.3972 [hep-th].}

\lref\sekino{
  B.~Freivogel, Y.~Sekino, L.~Susskind and C.~P.~Yeh,
  ``A holographic framework for eternal inflation,''
  Phys.\ Rev.\  D {\bf 74}, 086003 (2006)
  [arXiv:hep-th/0606204].
}

\lref\poppitz{
  E.~Ponton and E.~Poppitz,
  ``Gravity localization on string-like defects in codimension two and the
  AdS/CFT correspondence,''
  JHEP {\bf 0102}, 042 (2001)
  [arXiv:hep-th/0012033].
}

\lref\MaldacenaRE{
  J.~Maldacena,
  ``The large N limit of superconformal field theories and supergravity,''
  Adv.\ Theor.\ Math.\ Phys.\  {\bf 2}, 231 (1998)
  [Int.\ J.\ Theor.\ Phys.\  {\bf 38}, 1113 (1999)]
  [arXiv:hep-th/9711200].}

\lref\WittenQJ{  E.~Witten,
  ``Anti-de Sitter space and holography,''
  Adv.\ Theor.\ Math.\ Phys.\  {\bf 2}, 253 (1998)
  [arXiv:hep-th/9802150].
  }

\lref\GKP{
  S.~S.~Gubser, I.~R.~Klebanov and A.~M.~Polyakov,
  ``Gauge theory correlators from non-critical string theory,''
  Phys.\ Lett.\  B {\bf 428}, 105 (1998)
  [arXiv:hep-th/9802109].
}

\lref\doubletrace{
  E.~Witten,
  ``Multi-trace operators, boundary conditions, and AdS/CFT correspondence,''
  arXiv:hep-th/0112258;
  M.~Berkooz, A.~Sever and A.~Shomer,
  ``Double-trace deformations, boundary conditions and spacetime
  singularities,''
  JHEP {\bf 0205}, 034 (2002)
  [arXiv:hep-th/0112264].
}

\lref\weinberg{
  S.~Weinberg,
  ``Dynamics at infinite momentum,''
  Phys.\ Rev.\  {\bf 150}, 1313 (1966).
}

\lref\VanRaamsdonkFP{
  M.~Van Raamsdonk,
  arXiv:0802.3224 [hep-th].
}

\lref\Sakai{
  T.~Sakai and S.~Sugimoto,
``Low energy hadron physics in holographic QCD,''
  Prog.\ Theor.\ Phys.\  {\bf 113}, 843 (2005)
  [arXiv:hep-th/0412141].
}
  
\lref\SchaferIB{
  T.~Schafer,
  ``What atomic liquids can teach us about quark liquids,''
  Prog.\ Theor.\ Phys.\ Suppl.\  {\bf 168}, 303 (2007)
  [arXiv:hep-ph/0703141].
}
  
\lref\KovtunDE{
  A.~Buchel and J.~T.~Liu,
  ``Universality of the shear viscosity in supergravity,''
  Phys.\ Rev.\ Lett.\  {\bf 93}, 090602 (2004)
  [arXiv:hep-th/0311175];
  P.~Kovtun, D.~T.~Son and A.~O.~Starinets,
  ``Viscosity in strongly interacting quantum field theories from black hole
  physics,''
  Phys.\ Rev.\ Lett.\  {\bf 94}, 111601 (2005)
  [arXiv:hep-th/0405231].
}

\lref\superc{
  S.~S.~Gubser,
``Breaking an Abelian gauge symmetry near a black hole horizon,''
  arXiv:0801.2977 [hep-th];
  S.~S.~Gubser,
  ``Colorful horizons with charge in anti-de Sitter space,''
  arXiv:0803.3483 [hep-th];
  S.~A.~Hartnoll, C.~P.~Herzog and G.~T.~Horowitz,
  ``Building an AdS/CFT superconductor,''
  arXiv:0803.3295 [hep-th].
}

\lref\adsqcd{
  J.~Erlich, E.~Katz, D.~T.~Son and M.~A.~Stephanov,
  ``QCD and a holographic model of hadrons,''
  Phys.\ Rev.\ Lett.\  {\bf 95}, 261602 (2005)
  [arXiv:hep-ph/0501128].
}

\lref\unitarity{
    Y.~Nishida and D.~T.~Son,
  ``$\epsilon$ expansion for a Fermi gas at infinite scattering length,''
  Phys.\ Rev.\ Lett.\  {\bf 97}, 050403 (2006)
  [arXiv:cond-mat/0604500];
 F.~Werner and Y.~Castin,
  ``The unitary gas in an isotropic harmonic trap: symmetry properties 
  and applications,''
  Phys.\ Rev.\ {\bf A74}, 053604 (2006)
  [cond-mat/0607821];
     P.~Nikoli\'c and S.~Sachdev,
  ``Renormalization group fixed points, universal phase diagram, and 
  $1/N$ expansion for quantum liquids with interactions near the 
  unitarity limit,''
  Phys.\ Rev.\  {\bf A75}, 033608 (2007)
  [arXiv:cond-mat/0609106];
   M.~Y.~Veillette, D.~E.~Sheehy, and L~Radzihovsky,
  ``Large-$N$ expansion for unitary superfluid Fermi gases,''
  Phys.\ Rev.\ {\bf A75}, 043614 (2007)
  [arXiv:cond-mat/0610798];
  Y.~Nishida, D.~T.~Son and S.~Tan,
  Phys.\ Rev.\ Lett.\  {\bf 100}, 090405 (2008)
  [arXiv:0711.1562 [cond-mat.other]];
  T.~Mehen,
  arXiv:0712.0867 [cond-mat.other].
}

\lref\symmrefs{
 U.~Niederer,
  ``The maximal kinematical invariance group of the free Schr\"odinger
   equation,''
  Helv.\ Phys.\ Acta {\bf 45}, 802 (1972);
  C.~R.~Hagen,
"Scale and Conformal Transformations in Galilean-Covariant Field Theory,''
Phys.\ Rev.\ {\bf D5}, 377 (1972);
 R.~Jackiw and S.~Pi, ``Finite and Infinite Symmetries in (2+1)-Dimensional Field Theory," hep-th/9206092; 
}

\lref\handh{
P.~C.~Hohenberg and B.~I.~Halperin,
Rev.\ Mod.\ Phys.\ {\bf 49} 436 (1977).
}

\lref\expxi{
J.~T.~Stewart \etal, Phys.\ Rev.\ Lett.\ {\bf 97}, 220406 (2006)
[arXiv:cond-mat/0607776].}

\lref\Horvathy{
  C.~Duval, G.~W.~Gibbons and P.~Horvathy,
  ``Celestial Mechanics, Conformal Structures, and Gravitational Waves,''
  Phys.\ Rev.\  D {\bf 43}, 3907 (1991)
  [arXiv:hep-th/0512188].
}
\lref\alish{
  M.~Alishahiha, Y.~Oz and J.~G.~Russo,
  ``Supergravity and light-like non-commutativity,''
  JHEP {\bf 0009}, 002 (2000)
  [arXiv:hep-th/0007215].
}
\lref\mukund{
  V.~E.~Hubeny, M.~Rangamani and S.~F.~Ross,
  ``Causal structures and holography,''
  JHEP {\bf 0507}, 037 (2005)
  [arXiv:hep-th/0504034].
}

\Title{\vbox{\baselineskip12pt\hbox{0804.4053}
\hbox{MIT-CTP/3944}}}
{\vbox{
\centerline{Gravity duals for non-relativistic CFTs}}}
\bigskip
\centerline{Koushik Balasubramanian and John McGreevy}
\bigskip
\centerline{
{\it Center for Theoretical Physics,
Massachusetts Institute of Technology, Cambridge, MA 02139}} 
\bigskip
\bigskip
\noindent
We attempt to generalize the AdS/CFT correspondence
to non-relativistic conformal field theories
which are invariant under Galilean transformations.
Such systems govern ultracold atoms at unitarity,
nucleon scattering in some channels, and 
more generally, a family of universality classes
of quantum critical behavior.
We construct a family of metrics which realize these 
symmetries as isometries.
They are solutions of gravity with negative cosmological
constant coupled to pressureless dust.
We discuss realizations of the dust, which include
a bulk superconductor.
We develop the holographic dictionary
and find two-point correlators of the correct form.
A strange aspect of the correspondence
is that the bulk geometry has two extra noncompact dimensions.

\bigskip
\Date{April, 2008}

\newsec{Introduction}

Many attempts have been made to use 
the AdS/CFT correspondence \refs{\MaldacenaRE,\GKP,\WittenQJ}
to study 
systems realizable in a laboratory.
One does not yet have a 
holographic dual matching the precise microscopic 
details of any such system
and is therefore led to try to match
the universality class of the system.
In general, physics in the far infrared 
is described by a (sometimes trivial) 
fixed point of the renormalization group.
It has been argued that the 
associated zero-temperature conformal field theory (CFT)
controls a swath of the finite-temperature phase diagram, 
namely the region in which the temperature is the only important scale,
and no dimensionful couplings have turned on
(see \eg\ \sachdev).
In trying to use gauge/string duality to 
study such questions,
it seems urgent, then, to match the symmetries 
to those of the sytem of interest.

The AdS/CFT correspondence so far
gives an effective description of relativistic conformal field theories
at strong coupling.
Not many of these are accessible experimentally.

However, there are many {\it non-relativistic} conformal field
theories which govern physical systems.
Such examples arise in 
condensed matter physics \sachdev, atomic physics \zw, and nuclear physics 
\MehenND.
In the first situation,
these are called `quantum critical points';
this term refers to a second-order phase transition that is reached
not by varying the temperature, but by varying some coupling constants at zero temperature.

A particularly interesting example where exquisite experimental
control is possible is the case of cold fermionic atoms at unitarity (\eg\ \zw\ 
and references therein and thereto).
These are scale invariant in the following sense.
The interactions are tuned 
(by manipulating the energy of a two-body boundstate
to threshold)
to make the scattering length infinite (hence
the system is strongly coupled).
The material is dilute enough that the effective range 
of the potential can be treated as zero.
While the 
mass of the atoms
is a dimensionful quantity,
the dependence of the energy and other physical quantities
on that mass is determined by the symmetry algebra.
A recent paper studying 
the constraints from Galilean conformal symmetry on
such systems is \NishidaPJ\ (other recent theory references include
\unitarity).

In this brief note, we set out to find a bulk dual of non-relativistic CFTs,
analogous to the AdS gravity description of relativistic CFTs,
at strong coupling.
We approach this question by considering
the algebra of generators of the non-relativistic conformal group,
which appears in \NishidaPJ\
(related work includes \symmrefs).
We seek a solution of string theory (in its low-energy gravity approximation)
whose asymptotic boundary symmetry group is not the Poincar\'e symmetry group (\ie\ Lorentz and translations), but rather
Galilean invariance and translations.
We also demand a non-relativistic version of scale invariance 
and, when possible, special conformal transformations.

Clearly we will have to introduce some background energy density
to find such a solution (\ie\ it will not be just a solution of gravity with a cosmological constant).
Holographically, this is because the theories we are trying to describe
arise in general (perhaps not always)
from relativistic microscopic theories (not necessarily conformal)
upon by adding in some background of Stuff
(which fixes a preferred reference frame) and then taking some limit
focussing on excitations with small velocities in this frame.

An analogy which may be useful is the following.
The plane-wave symmetry group is a contraction of the AdS isometry group;
the plane wave solution arises from AdS in considering 
a limit focussing on the worldline of a fast-moving particle.
So it might be possible to take some limit of 
AdS analogous to the plane wave limit to find our solution.
The limit involved would be like the Penrose limit,
but replacing the single particle worldline with
a uniform background number density. 
We leave such a derivation from a 
more microscopic system for the future
and proceed to guess the end result.

Scale invariance can be realized 
in non-relativistic theories in a number of ways. 
One freedom is the relative scale dimension 
of time and space, called the `dynamical exponent' $z$
(see \eg\ \refs{\handh,\ArdonneWA}).  
The familiar Schr\"odinger case has $z=2$,
and the conformal algebra for this case is described in
\NishidaPJ.
Fixed points with $z \neq 2$ can be Galilean invariant, too.
We will find a metric with this symmetry group 
as its isometry group
for each value of $z$.\foot{
Gravity duals of related but distinct Lifshitz points (with time reversal
invariance) are being studied
by S.\ Kachru, X.\ Liu, and M.\ Mulligan \KLM.
We thank them for discussions.}
To be specific, the relevant algebra is:
\eqn\confalg{
\eqalign{
  & 
  [M_{ij},\, N] = 
 [M_{ij},\, D] = 0,
  ~~~ [M_{ij},\, P_k] = i(\delta_{ik} P_j - \delta_{jk} P_i), 
  ~~~  [M_{ij},\, K_k] = i(\delta_{ik} K_j - \delta_{jk} K_i) \cr
  & [M_{ij},\, M_{kl}]= i(\delta_{ik} M_{jk} - \delta_{jk} M_{il}
    + \delta_{il} M_{kj} - \delta_{jl} M_{ki}) \cr
      & [P_i,\, P_j] = [K_i,\, K_j] = 0, ~~~
    [K_i,\, P_j] = i \delta_{ij}N,~~~
 [D,\, P_i] = iP_i, ~~~ [D,\, K_i] = (1-z) i K_i  \cr
 &       [H,\, N] = [H,\, P_i] = [H,\, M_{ij}] = 0, 
~~  [H,\, K_i] = -i P_i, ~~ [D,\, H] = z iH, ~~
[D, \, N] = i (2-z) N
.}}
Here $i,j = 1..d$ label the spatial dimensions.
$M_{ij}$ generate spatial rotations, 
and the first two lines just state the properties of the various 
generators under spatial rotations.
$P_i$ are momenta, $K_i$ generate Galilean boosts,
$N$ is a conserved rest mass or particle number, and $D$
is the dilatation operator. 
That the particle number can be conserved is 
a consequence of the possibility of absence of antiparticles
in a non-relativistic theory\foot{That is to say,
the dual field theories that we are describing must not 
have particle production.  
It is of course possible for non-relativistic theories
to nevertheless contain both particles and antiparticles,
as indeed do the systems under study in \KLM,
which, unlike ours, are invariant under $ t \to - t$.}.

For the special case $z=2$, 
there is an additional special conformal generator $C$
which enjoys the algebra:
$$ 
 [M_{ij},\, C] = 0, \qquad
 [K_i,\, C] = 0, \qquad
     [D,\, C] = -2iC, \qquad
[H,\, C] = -i D .$$
The metrics we find are also invariant under the 
combined operations of time-reversal and charge conjugation.

%

%

%


In the next section we write down a metric with this isometry group.\foot{
Earlier work on geometric realizations of the Schr\"odinger group
include \Horvathy\ and references therein, wherein related metrics
are called Bargmann-Einstein structures.}
It is sourced by a stress tensor describing a negative
cosmological constant plus dust.
After demonstrating the action of the isometry group,
we provide answers to the 
important question: why doesn't the dust collapse due to the gravitational 
attraction?
In section three, we generalize the AdS/CFT dictionary 
to extract information about the field theory dual,
including anomalous dimensions and Green's functions.
In section four, we highlight some of the many open questions
raised by this construction.

\newsec{Geometry}

After all this talk, here, finally is the metric:
\eqn\zansatz{
ds^2 = L^2\(  -\frac{dt^2}{r^{2z}}  + \frac{ d \vec x^2 + 2d\xi dt}{r^2} + \frac{dr^2}{r^2}  \).
}
Here $\vec x$ is a $d$-vector,
and $z$ is the advertised dynamical exponent.
The coordinate $\xi$ is something new whose
interpretation we will develop below.

The metric \zansatz\
is nonsingular.  As is manifest in the 
chosen coordinates, it is conformal to a pp-wave spacetime\foot{
Note added: In the case $d=2, z=3$, this metric has appeared 
previously as the near-horizon
limit of D3-branes in the presence of lightlike tensor fields \alish;
its causal structure was studied in \mukund.  
The sources 
present in those solutions do not preserve the rotations $M_{ij}$.}.

\subsec{Symmetries}

We shall show that the isometries of the above metric 
comprise
the $d+1$-dimensional 
non-relativistic dilatation group
with algebra \confalg. It is clear that the above metric is invariant under translations in $\vec x, t$ and under rotations of $\vec x$. 
It is also invariant under the scale transformation
$$ x^{\prime} = \lambda x, ~~~
t^{\prime} = \lambda^{z}t, ~~~
r^{\prime} = \lambda r,~~~
\xi^{\prime} = \lambda^{2-z}\xi~~. $$

It is invariant under the Galilean boost
\eqn\gerbil{
\vec {x}^{ \prime} =\vec{x} - \vec {v}t
}
if we assign the following transformation to the coordinate $\xi$:
\eqn\gerbil{
\xi^{ \prime} = \xi + \frac{1}{2}(2 \vec{v} \cdot \vec{x} - {v^2}t)~~.
}
In the special case $z=2$, 
special conformal transformations act as 
$$
\vec x^{ \prime} = \frac{\vec x}{1+ct}, ~~~~
t^{ \prime}  = \frac{t}{1+ct}, ~~~~
r^\prime = {r \over {1+ct}}, ~~~~
\xi^{ \prime} = \xi + {c \over 2}{\vec{x}\cdot\vec{x} + r^2 \over 1 +  ct} ~~.
$$
To understand the transformation rule for $\xi$, 
let us define the transformations $T_1$ and $T_2$, such that
\eqn\gerbil{
T_1:(x,t) \rightarrow \(\frac{x}{t},\frac{1}{t}\),
~~~~~~~
T_2:(x,t) \rightarrow (x,t+c) ~~;
}
$T_1$ is an inversion and 
$T_2$ is a time-translation.
The special conformal transformation $S$ can be written as
$S = T_1T_2T_1.$
 
One can check that the vector fields generating
these isometries have Lie brackets
satisfying the algebra \confalg.\foot{We believe that 
no $d+1$-dimensional metric can realize this isometry group.}
The only non-obvious identification is 
that the rest mass is generated by
$ N = i\del_\xi$.  Note that although the metric is not invariant under
simple time reversal, $t\to - t$,
it is invariant under the combined operation
\eqn\discretesym{ t \to - t, ~~~ \xi \to - \xi ,}
which, given the interpretation of the
$\xi$-momentum as rest mass,
can be interpreted as the composition, $\CC \CT$,
of charge conjugation and time-reversal.

\subsec{Einstein's equations}

The stress tensor sourcing
the metric in equation \zansatz\ 
consists of a negative cosmological constant\foot{We have set $8\pi G_N=1$.
The indices $a,b$ run over all $d+3$ dimensions of the bulk.}
$\Lambda = -10$ plus 
a pressureless `dust', with constant density $\CE$:
$$ T_{ab} = - \Lambda g_{ab} -\CE \delta_{a}^0\delta_b^0 g_{00}. $$
For $d=3$, we find that the density is $ \CE = \( 2z^2  + z - 3 \)L^{-2}$.

In the following subsection, we consider what matter can produce
this stress tensor.
Our answers to this question do not
affect our later calculations, except 
to suggest what fields should be present in the bulk,
and to demonstrate that the metric \zansatz\ is physically sensible.

\subsec{Dust}

%

We observe that 
because of the non-stationary form of the metric,
the stress tensor for 
an electric field in the radial direction $F_{rt}$ 
has only a $00$ component.
Maxwell's equation in the bulk, coupled to a
background current $ j^a$ is
$$ {1\over \sqrt g} \del_a \( \sqrt g F^{ab}\) = j^b .$$
To produce an electric field in the $r$ direction, 
we consider a nonzero $j^\xi$ of the form
\eqn\jxi{ j^\xi = \rho_0 r^{\alpha}. }
Gauss' law implies that the electrostatic potential is
\eqn\vecpot{ A_0(r)  = 
{\rho_0 \over 
(\alpha-2) (\alpha-d-2) } r^{ \alpha - 2} .}
The current and gauge field are therefore related by the London equation
$ j_a = m_A^2 A_a$, with $m_A^2 = {z ( z+d) \over L^2}. $\foot{Recent holographic studies 
of 
spontaneously broken 
global symmetries of the boundary theory
include \adsqcd\
and
(at finite temperature)
\superc.}

Having gained this insight, we observe that 
our metric is sourced by the ground state of an Abelian Higgs model 
in its broken phase.  The model
$$
S = \int d^{d+3} x \sqrt g 
\( -{1\over 4} F^2 + {1\over 2} | D\Phi  |^2 - V\(|\Phi|^2\) \)  $$
with $ D_a\Phi \equiv (\del_a + i e A_a ) \Phi$,
with a Mexican-hat potential 
$$ V\(|\Phi|^2\) = g \( |\Phi|^2 - v^2 \)^2 + \Lambda $$
produces the correct dust stress tensor
for arbitrary $g$, as long as $e^2v^2 = m_A^2 = {z(z+d) \over L^2}$ as above 
\foot{Our metric is therefore a solution of the same 
system studied in the first and last references in \superc,
with very different asymptotics.}.
The parameter $\rho_0$ in the solution for the gauge field is determined by the Einstein equation to be
$\rho_0^{2} = \CE {z (z+d)^2\over 2z + d }L^4 $.
It is tempting to relate the 
phase of $\Phi$ to that of the boundary wavefunction.
We hope in the future to use this theory with finite $g$
to describe the vortex lattice formed by rotating
the cold-atom superfluids.

The fact that the metric \zansatz\ is sourced by a 
reasonable class of physical systems with stable ground states
suggests that it need have no intrinsic instabilities.

\newsec{Correspondence}

Consider a scalar field in the background \zansatz,
with action
$$ S = - \half \int d^{d+3}x \sqrt g \( \del_\mu \phi \del_\nu \phi g^{\mu\nu} 
+  m^2 \phi^2 \) ~ ,$$
Here $ d^{d+3}x\equiv d^d\vec x dt d\xi dr$.
The usual argument 
\GKP\
that $h_{xy}(k_z)$ 
satisfies the same equation of motion 
as a massless scalar goes through,
and so we can take the $m=0$ case as a proxy for 
this component of the metric fluctuations.
In the metric \zansatz, we have $ \sqrt g = r^{- (d+3)}$ for any $z$.
The wave equation in this background
has the form
$$ 0 = { 1\over \sqrt g} \del_\mu \( \sqrt g g^{\mu \nu} \del_\nu \phi \)- m^2 \phi 
= \[r^{d+3}\del_r \( {1\over r^{d+1}} \del_r \) + r^2 \( 2 \del_\xi \del_t 
+ r^{ 2 - 2 z} \del_\xi^2 + \vec \nabla^2\) - m^2 \] \phi.
$$

Translation invariance allows us to Fourier decompose
in the directions other than $r$,
$ \phi(r) = e^{ i \omega t + i \vec k \cdot x + i l \xi } f_{\omega, \vec k, l}(r).$
Then: 
$$  \[- r^{d+3} \del_r \( {1\over r^{d+1}} \del_r \) + r^2 (  2 l \omega + \vec k^2 )
+ r^{ 4 - 2z} l^2 + m^2\] f_{\omega, \vec k, l}(r) = 0. $$
To find the generalization of the gauge/gravity dictionary
to this case, we first study asymptotic solutions of this equation near the boundary.
Writing $ f \propto r^{\Delta}$, we find for $z \leq 2$:
\eqn\critexptwo{\Delta_\pm =  
1 +{d\over 2}\pm 
\sqrt{ \(1 + {d\over 2}\)^2 + m^2 + \delta_{z, 2} l^2 }
~~~~.}
In the case $z > 2$, there is no scaling solution near the boundary;
the $\xi$-momentum dominates the boundary behavior,
and the solutions vary with $r$ faster than any power of $r$.
It would be interesting to understand
a possible meaning of the asymptotic solutions which do arise,
and to interpret the critical nature of $z=2$ in terms of the boundary theory.
As usual, these exponents $\Delta_\pm$ can be interpreted
as scaling dimensions of spin-zero boundary operators 
in the boundary theory.
We look forward to extending 
this analysis to the fluctuations of fields of
other spin.
Instead we will proceed to 
compute correlation functions of these scalar operators.

\subsec{Correlators}

For definiteness, we focus 
on the familiar, critical $z=2$ case with $d=3$ spatial dimensions.
In that case, the momentum in the $\xi$-direction
simply adds to the mass of the bulk scalar, \ie\
they appear in the combination $ l^2 + m^2 $.
In this case, the wave equation becomes
$$  \[- r^6 \del_r \( {1\over r^4} \del_r \) + r^2 (  2 l \omega + \vec k^2 )
+ l^2 + m^2\] f_{\omega, \vec k, l}(r) = 0 $$
and is solved by
$$ f_{\omega, \vec k, l}(r) = A r^{5/2} K_\nu( \kappa r) $$
where $K$ is a modified Bessel function, 
$A$ is a normalization constant, and\foot{
For general $d$, with $z= 2$, the solution is
$$ f_{\omega, \vec k, l}(r) = A r^{1 + {d\over 2}} K_\nu( \kappa r) $$
with 
$$ \nu = \sqrt{ \(1 + {d\over 2}\)^2 + M^2 }, ~~~ \kappa^2  = 2 l \omega+ \vec k^2, 
~~~M^2 \equiv m^2 + \delta_{z,2} l^2  .$$
} 
$$ \nu = \sqrt{ \( {5\over 2}\)^2  + l^2 + m^2} , ~~~ \kappa^2  = 2 l \omega+ \vec k^2  .
$$
As usual in AdS/CFT, we choose $K$ over $I$ because
it is well-behaved near the horizon, $K(x) \sim e^{ - x }$ at large $x$.

To compute correlators, we 
introduce a cutoff near the boundary at $r = \epsilon$,
and normalize $f_\kappa(\epsilon) = 1$
(so $A = \epsilon^{-5/2}K_\nu(\kappa \epsilon)^{-1}$).
The usual prescription goes through,
leaving an on-shell bulk action of the form
$$  S[\phi_0] 
= \half \[ \int d^{d+2}X \sqrt g g^{rr} \phi(X) \del_r \phi(X) \]_{r= \epsilon}
$$
where $X \in \{\vec x, t, \xi\}$ are the coordinates on the boundary.
This evaluates to 
$$ S[\phi_0] 
=\half \int dp \phi_0(-p) \CF(\kappa, \epsilon) \phi_0(p) 
$$
with the `flux factor'
$$ \CF(\kappa, \epsilon) = 
\lim_{r \to \epsilon} ~\sqrt g g^{rr} f_\kappa(r) \del_r f_\kappa(r)=
\sqrt g g^{rr} \del_r \ln \( r^{{2 + d \over 2}}K_\nu(\kappa r)\)
|_{r= \epsilon}~~.$$
Using the expansion for small $x$
$$ K_\nu(x) \simeq 2^{ \nu -1 } \Gamma(\nu) x^{ - \nu} 
\( 1 + ... - \left( {x\over 2}\right)^{2 \nu} {\Gamma(1-\nu) \over \Gamma(1+\nu)} \)$$
and ignoring contact terms, this is 
$$ \CF(\kappa, \epsilon) = 
- { \Gamma(1-\nu) \over \Gamma(\nu) } {1\over 2 \epsilon^5} 
\left( { \kappa \epsilon \over 2 } \right)^{2 \nu}. $$
This gives
$$ 
\vev{ \CO_1(\omega, \vec k) \CO_2(\omega', \vec k') } 
\propto   \delta( k + k') {\Gamma(1-\nu) \over \Gamma(\nu)}
{1\over 2  \epsilon^{5}}
\left( {  (  2 l  \omega + \vec k^2 ) \epsilon^2 \over 4 } \right)^{ \nu}
$$
(in the case $\phi = h_{xy}$, $\CO = T_{xy}$).
We defer a more careful treatment of holographic renormalization to future work.
Fourier transforming to position space, this is
$$ \vev{ \CO_1(x, t) \CO_2(0,0) } 
\propto
{\Gamma(1-\nu) \over \Gamma(\nu)}
\delta_{\Delta_1, \Delta_2}
\theta(t) 
 { 1\over  |\epsilon^2 t|^{\Delta} } e^{ - i l x^2 / 2 |t|}
$$
where we used translation invariance to put the second operator at $(\vec x,t) = (\vec 0,0)$, $\theta$ is a step function,
and $\Delta_{i,j}$ are the scaling dimensions of the operators
that we found in \critexptwo.
This is the form one expects for the correlators
in a Galilean-invariant conformal field theory in $d$ spatial dimensions
with dynamical exponent $2$.  
It will be interesting to extend this calculation to other values of $z$.


\newsec{Discussion}

In discussions of AdS/CFT, one often hears that the CFT 
`lives at the boundary' of the bulk spacetime.
The spacetime \zansatz\ is 
conformal to a pp-wave spacetime,
and hence has a boundary which is one dimensional --
for $z>1$, $g_{tt}$ grows faster than the other components
at small $r$.
While one might be tempted to speculate
that this means that the dual field theory is 
one dimensional, there is no clear evidence for this statement
(in the usual case, only the UV of the field theory 
can really be said to live at the boundary of the AdS space),
and indeed we have nevertheless used it to compute
correlators which look like those of a $d$-dimensional nonrelativistic
field theory.

Perhaps the most mysterious aspect of this new correspondence
is not of too few dimensions, but of too many\foot{
Other codimension-two holographic relations include 
\sekino, where one of the missing dimensions is time,
and \poppitz, which can be understood by a 
Kaluza-Klein decomposition with discrete momenta.}.
The $\xi$ direction is an additional noncompact dimension
of the bulk geometry besides the usual radial direction,
which seems to be associated to the conserved rest mass\foot{
The similarity with the plane wave appears here, too: 
the plane wave describes a relativistic system
in the infinite momentum frame, where there is also 
no particle production \weinberg.
We thank Allan Adams for this observation.
}.
We have computed correlators at fixed $l$;
this is reasonable since the theory is 
non-relativistic and so a sector with definite $l$ 
can be closed (unlike in a relativistic theory
where these sectors unavoidably mix by production of 
back-to-back particles and antiparticles).
At the special value of the dynamical exponent $z=2$,
varying $l$ affects not just the units in which 
energy is measured, but 
the values of the critical exponents (see equation \critexptwo).

We would like to think of these solutions 
as scaling limits which zoom in on 
small fluctuations about particular
states of relativistic theories (possibly CFTs),
where a preferred rest frame is fixed by 
some density of Stuff.  
It would be nice to see our metric arise as such a contraction of 
an asymptotically AdS solution.
Such an identification
may help find the finite-temperature geometry.

A word of caution in this direction is in order:
at large R-charge but low temperature,
there is an instability in supersymmetric theories
to Bose condensation, because there are inevitably scalar fields
that carry R-charge.  A system with symmetries
under which only fermions are charged (such as \Sakai) could be 
useful, and would seem more likely to be related to 
cold fermions away from the fixed point.

\bigskip

Many questions remain.
We have computed the correlator $ \vev{T_{xy}T_{xy}}$.
Does this plus the dynamical exponent completely
determine the $TT$ correlator, or
are there independent central charges 
in the non-relativistic theory?

One interesting observable,
which has been {\it measured} as well as computed
numerically for cold-fermion systems
is 
the analog of the famous $3/4$ in the $\CN=4$ SYM theory, 
namely
the ratio of the energy density at unitarity
to the energy density of the corresponding free theory
($\xi$ in \SchaferIB\ and \unitarity, proportional to $\beta$ in 
\expxi).\foot{We thank
Krishna Rajagopal for teaching us about this.}
We will need to learn to compute the analog of the ADM
mass for spacetimes with the asymptopium of \zansatz.

Once the finite-temperature solution with
these asymptotics is found,
there will be many more interesting things to compute.
Of course, it will be nice to check the universality of $ \eta \over s$
\KovtunDE, obtainable from the correlator we have studied, at nonzero temperature.
As a possible but quite ambitious example of a qualitative consequence
of the non-relativistic microscopic constituents
in these systems we offer the following.
It has been noted \refs{\eg\ \VanRaamsdonkFP} that
the plasma made by putting
any relativistic CFT at finite temperature
is quite compressible.  The compressibility
is known to have an important
effect on the turbulence cascade 
\cascade; it can even change its direction.
It is therefore of interest to have gravity descriptions
of less compressible fluids,
which these would presumably be.

\bigskip
While this work was being completed, we learned that 
Dam Son has independently found very closely related results \damson.

\bigskip
\centerline{\bf{Acknowledgements}}
We thank 
Allan Adams,
Shamit Kachru, Pavel Kovtun, Hong Liu, Krishna Rajagopal,
Mukund Rangamani, 
Brian Swingle 
and Alessandro Tomasiello
for discussions and encouragement.
Thanks to Joe Minahan for checking our algebra.

This work
was supported in part by funds provided by the U.S. Department of
Energy (D.O.E.) under cooperative research agreement
DE-FG0205ER41360.

\listrefs
\end